\documentclass{article} \usepackage{amssymb,latexsym,amsfonts} 
\usepackage{fullpage}
\usepackage{psfig} 
\title{An Exactly Solvable Model For The
Interaction Of Linear Waves With Korteweg-de Vries Solitons}
\author{P. D. Miller\thanks{Current affiliation: Department of
Mathematics, University of Michigan, East Hall, 525 E. University
Avenue, Ann Arbor, MI 48109.  Phone: (734) 647-4473.  Fax: (734)
763-0937.  Email: {\tt millerpd@umich.edu}}\,\, and
S. R. Clarke\thanks{Phone: +61-3-9905-4421.  Fax: +61-3-9905-4403.
Email: {\tt src@mail.maths.monash.edu.au}}\\ Department of Mathematics
and Statistics\\ Monash University\\ P. O. Box 28M, Victoria 3800\\
Australia} \date{January 16, 2001\\ To appear in {\em SIAM
J. Math. Anal.}}  \newtheorem{theorem}{Theorem}
\newcommand{\pfend}{$\Box$}
\newcommand{\res}[1]{\begin{array}{c}\\{\rm
Res}\\\scriptstyle{#1}\end{array}}
\begin{document}
\maketitle
\begin{abstract}
Under certain mode-matching conditions, small-amplitude waves can be
trapped by coupling to solitons of nonlinear fields.  We present a
model for this phenomenon, consisting of a linear equation coupled to
the Korteweg-de Vries (KdV) equation.  The model has one parameter, a
coupling constant $\kappa$.  For one value of the coupling constant,
$\kappa=1$, the linear equation becomes the linearized KdV equation,
for which the linear waves can indeed be trapped by solitons, and
moreover for which the initial value problem for the linear waves has
been solved exactly by Sachs \cite{S83} in terms of quadratic forms in
the Jost eigenfunctions of the associated Schr\"odinger operator.  We
consider in detail a different case of weaker coupling, $\kappa=1/2$.
We show that in this case linear waves may again be trapped by
solitons, and like the stronger coupling case $\kappa=1$, the initial
value problem for the linear waves can also be solved exactly, this
time in terms of linear forms in the Jost eigenfunctions.  We present
a family of exact solutions, and we develop the completeness relation
for this family of exact solutions, finally giving the solution
formula for the initial value problem.  For $\kappa=1/2$, the
scattering theory of linear waves trapped by solitons is developed.
We show that there exists an explicit increasing sequence of
bifurcation values of the coupling constant, $\kappa=1/2,1,5/3,\dots$,
for which some linear waves may become trapped by solitons. By
studying a third-order eigenvalue equation, we show that for
$\kappa<1/2$ all linear waves are scattered by solitons, and that for
$1/2<\kappa<1$ as well as for $\kappa>1$ some linear waves are {\em
amplified} by solitons.
\end{abstract}

\section{Introduction}
This paper is concerned with solving the coupled system of equations:
\begin{eqnarray}
\displaystyle \partial_tA + \partial_x\left[\frac{1}{2}A^2 + \partial_x^2A\right] &=&0
\label{eq:kdv}\\
\nonumber\\
\displaystyle
\partial_tB +\partial_x\left[\kappa AB + \partial_x^2B\right] &=&0\,,
\label{eq:linear}
\end{eqnarray}
where $\kappa$ is a real parameter.  Of course, the nonlinear equation
for $A(x,t)$ is simply the Korteweg-de Vries equation, and it can be
solved independently by the inverse-scattering transform \cite{GGKM67}.
The coupled system (\ref{eq:kdv}) and (\ref{eq:linear}) is a partially
linearized version of the system proposed by Hirota and Satsuma \cite{HS81}
as a model for the dynamics of coupled long waves.  

The coupled system (\ref{eq:kdv}) and (\ref{eq:linear}) can be solved
exactly when $\kappa=1$ and when $\kappa=1/2$.  The case of $\kappa=1$
is well-known, for then the equation (\ref{eq:linear}) is just the KdV
equation itself linearized about the solution $A(x,t)$.  An elementary
exact solution of the linear equation (\ref{eq:linear}) in this case
is given by $B(x,t)=\partial_xA(x,t)$.  Further solutions can be
expressed in terms of derivatives of the squared eigenfunctions of the
related Schr\"odinger operator with potential $A$ \cite{S83}.

The case of $\kappa=1/2$ is essentially different.  In this case, the
linear equation (\ref{eq:linear}) is no longer the linearization of
KdV about {\em any} solution.  An elementary exact solution of the
linear equation in this case is given simply by $B(x,t)=A(x,t)$.  The
main goal of this paper is to construct the {\em general} solution of
the initial value problem for this linear equation when $A(x,t)$ is a
multisoliton solution of KdV.

\begin{figure}[h]
\begin{center}
 \mbox{\psfig{file=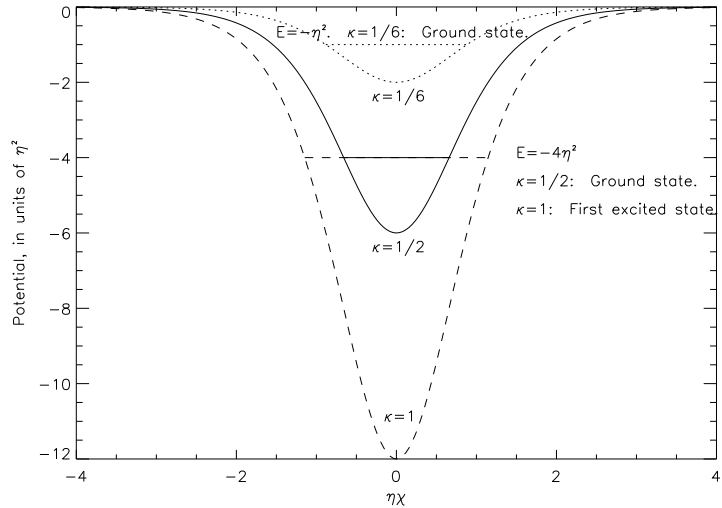,width=4 in}}
\end{center}
\caption{\em Energy levels of the $12\kappa \eta^2 {\rm sech}^2(\eta\chi)$
potential for three different values of the coupling constant
$\kappa$.}
\label{fig:wells}
\end{figure}

One way to make clear the difference between the cases $\kappa=1$ and
$\kappa=1/2$ is to consider $A(x,t)$ to be the simple soliton solution
of KdV (\ref{eq:kdv}):
\begin{equation}
A(x,t)=12\eta^2{\rm sech}^2(\eta (x-4\eta^2 t -\alpha)) = -V(\chi)\,,
\label{eq:simplesoliton}
\end{equation}
where $\chi=x-ct-\alpha$ and the velocity is $c=4\eta^2$.  If we look for
solutions of the linear equation (\ref{eq:linear}) that are traveling
waves with speed $c$, we find the equation
\begin{equation}
\left[-\kappa V(\chi)B(\chi) + B''(\chi)\right]' = cB'(\chi)\,.
\end{equation}
Integrating once, using vanishing boundary conditions at $\chi=\pm\infty$,
yields a Schr\"odinger eigenvalue problem for $B$:
\begin{equation}
-B''(\chi) +\kappa V(\chi)B(\chi) = E B(\chi)\,,
\end{equation}
where $E=-c$.  For $\kappa=1/2$, it follows from the fact that
$B(x,t)=A(x,t)$ is a solution of (\ref{eq:linear}) that the function
$B(\chi)=V(\chi)$ is an eigenfunction of the Schr\"odinger operator
with eigenvalue $E=-c=-4\eta^2$.  Since it has no zeros, it is the
ground state eigenfunction.  We will see below that there is also one
excited state for $\kappa=1/2$, although it is not relevant here since
it corresponds to a different velocity.  On the other hand, for
$\kappa=1$, $B(x,t)=\partial_x A(x,t)$ is a solution of
(\ref{eq:linear}), which implies that the function $B(\chi)=\partial_x
V(\chi)$ is an eigenfunction of the Schr\"odinger operator with the
same eigenvalue $E=-c=-4\eta^2$.  In this case, the eigenfunction has
a single zero and therefore is the first excited state.  It follows
that there are at least two eigenvalues for $\kappa=1$.  In fact,
there are exactly three states in this case.  A final observation is
that from the construction of the one-soliton solution of KdV (see
below) it follows that for $\kappa=1/6$, the function
$B(\chi)=V(\chi)$ is an eigenfunction of the Schr\"odinger operator
with eigenvalue $E=-c/4=-\eta^2$.  It is the ground state, and the
only eigenfunction.  These relationships are summarized in
Figure~\ref{fig:wells}.
We will have more to say about this picture when we discuss the
trapping of linear waves by solitons for general values of $\kappa$ in
\S~\ref{sec:numerics}.

The rest of this paper is primarily concerned with developing the
general solution of the initial value problem for (\ref{eq:linear})
with $\kappa=1/2$ when $A(x,t)$ is an $N$-soliton solution of KdV
(\ref{eq:kdv}).  In \S~\ref{sec:solutionformulas} we show how for
$\kappa=1/2$ a large family of exact solutions of (\ref{eq:linear})
can be obtained from the simultaneous solutions of the Lax pair for
KdV.  When the solution of KdV contains only solitons and no
radiation, the construction of Lax eigenfunctions is completely
algorithmic and algebraic, and consequently the corresponding family
of solutions of (\ref{eq:linear}) for $\kappa=1/2$ can be obtained
with great practicality.  In \S~\ref{sec:completeness} we then
establish that in the $N$-soliton case there are enough of these exact
solutions of (\ref{eq:linear}) for $\kappa=1/2$ to expand for fixed
$t$ any absolutely continuous $L^1({\mathbb R})$ function of $x$.
This fact then leads to a general solution formula for
(\ref{eq:linear}) simply by expanding the initial data.  We present
and discuss this formula in \S~\ref{sec:general}.  There will turn out
to be $N$ linearly independent solutions that are asymptotically
confined to the union of soliton trajectories and can therefore be
considered to be bound states.  In \S~\ref{sec:scattering} we compute
the scattering matrix that relates the asymptotic behavior of bound
states for $t\rightarrow -\infty$ to the corresponding behavior for
$t\rightarrow+\infty$.  In \S~\ref{sec:numerics} we consider general
values of the coupling constant $\kappa$ and describe the behavior of
some solutions of (\ref{eq:linear}) when $A(x,t)$ is a one-soliton
solution of KdV (\ref{eq:kdv}).  These calculations indicate the
exceptional nature of the two values $\kappa=1/2$ and $\kappa=1$.  In
the Appendix, we describe several physical applications of the coupled
system (\ref{eq:kdv}) and (\ref{eq:linear}) to topics in molecular
dynamics, mechanics, soliton theory, and the fluid dynamics of
internal waves.

\section{Exact Solution Formulas for $\kappa=1/2$}
\label{sec:solutionformulas}
As is well-known \cite{GGKM67}, the KdV equation (\ref{eq:kdv}) is the
compatibility condition for a pair of linear equations involving a
complex parameter $\lambda$ for an auxiliary function
$f(x,t,\lambda)$.  This pair of linear equations is
\begin{equation}
\partial_x^2f = -\frac{\lambda^2}{4}f - \frac{1}{6} Af\,,\hspace{0.2 in}
\mbox{ and }\hspace{0.2 in}
\partial_t f=\frac{1}{6} \partial_x A\cdot f + \left(\lambda^2 -\frac{1}{3}A\right)\partial_x f\,,
\label{eq:lax}
\end{equation}
and is called a Lax pair.  A simultaneous solution $f(x,t,\lambda)$ of
these linear equations exists if and only if the function $A(x,t)$
satisfies KdV (\ref{eq:kdv}).  Suppose that this is the case.  Then,
it is a direct matter to verify that for fixed but arbitrary
$\lambda\in{\mathbb C}$, the two functions defined by
\begin{equation}
B(x,t):=\partial_x\left[f(x,t,\lambda)\exp\left(\pm\frac{i}{2}(\lambda x +
\lambda^3 t)\right)\right]\,,
\label{eq:solutionformula}
\end{equation}
are solutions of the linear equation (\ref{eq:linear}) when
$\kappa=1/2$.  Note here an important point of departure from the
other solvable case, namely $\kappa=1$, where (\ref{eq:linear}) is the
linearized KdV equation.  In the latter case, particular solutions are
expressed in terms of the $x$-derivative of the {\em square} of the
Lax eigenfunction $f(x,t,\lambda)$ \cite{GGKM74,S83}.  By contrast,
the formula (\ref{eq:solutionformula}) for solutions of
(\ref{eq:linear}) for $\kappa=1/2$ is linear in $f(x,t,\lambda)$.
This fact leads to some important simplifications.

The formula (\ref{eq:solutionformula}) is only really practical to use
if one can explicitly compute the function $f(x,t,\lambda)$.  This
will be the case if the solution $A(x,t)$ of KdV (\ref{eq:kdv}) is a
pure $N$-soliton solution.  For each fixed $t$, $A(x,t)$ is then a
reflectionless potential of the Schr\"odinger equation in the Lax pair
(\ref{eq:lax}).  The multisoliton solutions of KdV and the associated
solutions of the Lax pair are constructed as follows \cite{KM56}.  Let
$f_+(x,t,\lambda)$ be given by
\begin{equation}
f_+(x,t,\lambda):=\left(1+\sum_{n=0}^{N-1}\lambda^{n-N}f_n(x,t)\right)\exp\left(-\frac{i}{2}(
\lambda x +\lambda^3t)\right)\,,
\label{eq:formoff}
\end{equation}
where the $f_n(x,t)$ are unknown coefficients.  Choose $N$ positive
numbers $\eta_1>\eta_2>\dots>\eta_N$, and $N$ arbitrary real numbers
$\alpha_1,\dots,\alpha_N$, and insist that $f_+(x,t,\lambda)$ satisfy
the relations
\begin{equation}
f_+(x,t,2i\eta_n)=(-1)^{n+1}\exp(2\eta_n\alpha_n)f_+(x,t,-2i\eta_n)\,,
\label{eq:KMrelations}
\end{equation}
for all $n=1,\dots,N$.  It is easy to see that these relations imply a
square linear algebraic system for the coefficients $f_n(x,t)$.  The
determinant of the system is always nonzero, and so the coefficients
$f_n(x,t)$ are determined uniquely from the soliton eigenvalues
$\{\lambda_n=2i\eta_n\}$ and the norming constants $\{\alpha_n\}$ in
terms of exponential functions.  From this construction, it can be shown that
if one chooses
\begin{equation}
A(x,t):= 6i\partial_x f_{N-1}(x,t)\,,
\label{eq:kdvsolution}
\end{equation}
then $f_+(x,t,\lambda)$ and $f_-(x,t,\lambda):= f_+(x,t,-\lambda)$ are
two simultaneous solutions of the Lax pair (\ref{eq:lax}), and the
function $A(x,t)$ defined by (\ref{eq:kdvsolution}) satisfies KdV
(\ref{eq:kdv}).  The two functions $f_\pm(x,t,\lambda)$ are linearly
independent for all nonzero $\lambda\neq \pm 2i\eta_n$.  According to
the linear relations (\ref{eq:KMrelations}) that determine the
coefficients, at the exceptional values of $\lambda$ the two functions
are proportional.

The solution $A(x,t)$ of KdV so constructed represents the interaction
of $N$ solitons.  In particular, as $t\rightarrow \pm\infty$, the solution
can be represented in the form
\begin{equation}
A(x,t)\sim\sum_{n=1}^N A_n^\pm (x,t)\,,\hspace{0.2 in}\mbox{ where }
\hspace{0.2 in}A_n^\pm(x,t):= 12\eta_n^2{\rm
sech}^2(\eta_n(x-\alpha_n^\pm)-4\eta_n^3t)\,,
\label{eq:Andef}
\end{equation}
where the asymptotic phase constants $\alpha_n^\pm$ are functions of
the $\eta_n$ and $\alpha_n$.  

Below, we will need the asymptotic behavior of the functions
$f_\pm(x,t,\lambda)$ as $x\rightarrow \pm\infty$ for $\lambda$ and $t$
fixed.  It can be shown that the coefficient functions $f_n(x,t)$
remain bounded as $x\rightarrow\pm\infty$.  Then, letting $x$ tend to
$\pm\infty$ in the linear relations (\ref{eq:KMrelations}), one finds
from dominant balance arguments that
\begin{equation}
\lim_{x\rightarrow\pm\infty} \left(1+\sum_{n=0}^{N-1}\lambda^{n-N}f_n(x,t)\right)\Bigg|_{\lambda=\pm 2i\eta_n} = 0\,.
\end{equation}
These relations imply that
\begin{equation}
\lim_{x\rightarrow\pm\infty}\left(1+\sum_{n=0}^{N-1}(\pm\lambda)^{n-N}f_n(x,t)\right) = \lambda^{-N}\prod_{n=1}^N (\lambda-2i\eta_n)\,.
\end{equation}
Therefore, for all $\lambda\in\mathbb{C}$,
\begin{equation}
\lim_{x\rightarrow\pm\infty}f_+(x,t,\lambda)\exp\left(\frac{i}{2}(\lambda x +\lambda^3 t)\right) = \lambda^{-N}\prod_{n=1}^N(\lambda\mp 2i\eta_n)\,.
\end{equation}
The large $|x|$ asymptotics for the other solution $f_-(x,t,\lambda)$
follow from the definition $f_-(x,t,\lambda)=f_+(x,t,-\lambda)$.  From
these asymptotics, it is easy to construct the appropriate linear
combinations of $f_\pm(x,t,\lambda)$ that correspond to the Jost
functions of the Schr\"odinger equation, normalized at $x=\pm\infty$.

In the formula (\ref{eq:solutionformula}) we have a choice of sign in
the exponent.  In fact, it is easy to see that if one considers the
totality of solutions obtained for all complex $\lambda$, the choice
of sign is redundant.  In what follows, we adopt a particular choice
of the sign and maintain generality by using both Lax eigenfunctions
$f_+(x,t,\lambda)$ and $f_-(x,t,\lambda)$.  Thus, the particular
solutions of the linear equation (\ref{eq:linear}) for $\kappa=1/2$
that we will consider below will be denoted by $h_\pm(x,t,\lambda)$,
given by
\begin{equation}
h_\pm(x,t,\lambda) := \partial_x g_\pm(x,t,\lambda)\,,\hspace{0.2 in}
\mbox{ where }\hspace{0.2 in}
g_\pm(x,t,\lambda) := f_\pm(x,t,\lambda)\exp\left(\frac{i}{2}(\lambda x +
\lambda^3 t)\right)\,.
\label{eq:particularsolutions}
\end{equation}
From the Schr\"odinger equation (\ref{eq:lax}) for
$f_\pm(x,t,\lambda)$ it follows that $g_\pm(x,t,\lambda)$ satisfies
the ODE
\begin{equation}
-i\partial_x^2g_\pm -i\frac{A}{6}g_\pm = \lambda \partial_xg_\pm\,.
\label{eq:ODE}
\end{equation}
This ODE plays an important role in suggesting the completeness
relation for the solutions $h_\pm(x,t,\lambda)$.

\section{The Completeness Relation for $\kappa=1/2$}
\label{sec:completeness}
Having in hand a large family of exact solutions of the linear
equation (\ref{eq:linear}) for $\kappa=1/2$ is certainly useful, but
we may then ask whether there are enough of these solutions to
construct the general solution of the initial value problem by
superposition.  A {\em completeness relation} is a formula that gives
the expansion of arbitrary initial data in terms of such a collection
of functions.  In this section, we will establish the completeness
relation for the exact solutions $h_\pm(x,t,\lambda)$ obtained in
\S~\ref{sec:solutionformulas}.

The form of the completeness relation is suggested by a similar
argument to that used by Sachs \cite{S83} in his investigation of the
completeness of squared eigenfunction solutions to the linearized KdV
equation.  The idea is that, ideally we would like to have a
differential eigenvalue problem in standard form satisfied by the
functions $h_\pm(x,t,\lambda)$:
\begin{equation}
L(t)h_\pm(x,t,\lambda)=\lambda h_\pm(x,t,\lambda)\,,
\end{equation}
where $\lambda$ is the eigenvalue and $L(t)$ is some second-order
linear differential operator in $x$.  Then, using the two explicit
solutions $h_\pm(x,t,\lambda)$ of this problem, we could solve the
inhomogeneous problem
\begin{equation}
L(t)\psi-\lambda\psi = \phi
\label{eq:hinhomog}
\end{equation}
by variation of parameters, {\em i.e.} by writing $\psi$ as a linear
combination of $h_\pm(x,t,\lambda)$ with nonconstant coefficients, and
substituting into (\ref{eq:hinhomog}).  For a fixed function
$\phi(x)$, this determines $\psi(x,t,\lambda)$, and we have thus
constructed the resolvent of the operator $L(t)$,
\begin{equation}
\psi(x,t,\lambda)=(L(t)-\lambda\mathbb{I})^{-1}\phi(x)\,.
\label{eq:resolvent}
\end{equation}
If the spectrum of $L(t)$ is contained in a bounded region of the complex
plane (and also under some milder conditions), then the Dunford-Taylor
integral of the resolvent on a positively oriented contour enclosing
the spectrum yields the identity operator:
\begin{equation}
-\frac{1}{2\pi i}\oint \psi(x,t,\lambda)\,d\lambda =-\frac{1}{2\pi
i}\oint (L(t)-\lambda\mathbb{I})^{-1}\phi(x)\,d\lambda =\phi(x)\,.
\label{eq:DunfordTaylor}
\end{equation}

But we do not have a second-order eigenvalue problem for
$h_\pm(x,t,\lambda)$.  Instead we have the second-order equation
(\ref{eq:ODE}) for $g_\pm(x,t,\lambda)$.  However, we make the {\em
guess} that a similar procedure will apply here.  Namely, for
appropriate side conditions (see below) we solve the inhomogeneous
equation
\begin{equation}
-i\partial_x^2\psi-i\frac{A}{6}\psi -\lambda\partial_x\psi = \phi\,,
\label{eq:ginhomog}
\end{equation}
for $\psi(x,t,\lambda)$ using variation of parameters with the two
functions $g_\pm(x,t,\lambda)$ solving the homogeneous equation
(\ref{eq:ODE}), and then we differentiate the resulting formula with
respect to $x$.  Formally speaking only, we have thus constructed the
resolvent of the ``operator''
\begin{equation}
L(t)=-i\partial_x -i\frac{A}{6}\partial_x^{-1}\,.
\label{eq:phonyoperator}
\end{equation}
The obstruction to rigor here is that $\partial_x^{-1}$ is not well-defined.
Nonetheless, we are guided to hypothesize that
\begin{equation}
-\frac{1}{2\pi i}\oint\partial_x\psi(x,t,\lambda)\,d\lambda = \phi(x)\,,
\label{eq:integralidentity}
\end{equation}
for an appropriate contour of integration.  This formula turns out to
be correct, although a direct proof must be supplied.  The proof we
use follows Miller and Akhmediev \cite{MA98}.

\subsection{Solving the inhomogeneous problem.}
We express the solution of the inhomogeneous problem in the form
\begin{equation}
\psi(x,t,\lambda) = C_+(x,t,\lambda)g_+(x,t,\lambda) + C_-(x,t,\lambda)g_-(x,t,\lambda)\,,
\label{eq:variationofparameters}
\end{equation}
subject to the usual ``reduction of order'' condition
\begin{equation}
\partial_xC_+(x,t,\lambda)\cdot g_+(x,t,\lambda) + \partial_xC_-(x,t,\lambda)
\cdot g_-(x,t,\lambda)=0\,.
\label{eq:simplify}
\end{equation}
Substituting (\ref{eq:variationofparameters}) into the equation for
$\psi$, and using (\ref{eq:simplify}), one finds
\begin{equation}
\partial_xC_+(x,t,\lambda)=-i\frac{\phi(x)g_-(x,t,\lambda)}{W(g_+,g_-)}\,,
\hspace{0.2 in}\mbox{ and }\hspace{0.2 in}
\partial_xC_-(x,t,\lambda)=i\frac{\phi(x)g_+(x,t,\lambda)}{W(g_+,g_-)}\,,
\label{eq:coefficients}
\end{equation}
where $W(g_+,g_-):= g_+\partial_xg_--g_-\partial_xg_+$ is the Wronskian.

From the differential equation (\ref{eq:ODE}) satisfied by
$g_\pm(x,t,\lambda)$, it follows that
\begin{equation}
\partial_x W(g_+,g_-)=i\lambda W(g_+,g_-)\,.  
\label{eq:WronskianODE}
\end{equation}
Using the large $|x|$ asymptotics of $f_\pm(x,t,\lambda)$ obtained in
\S~\ref{sec:solutionformulas} one then solves (\ref{eq:WronskianODE})
uniquely and finds that
\begin{equation}
W(g_+,g_-) = i\lambda^{1-2N}\exp(i(\lambda x +\lambda^3 t))\prod_{n=1}^N
(\lambda^2 + 4\eta_n^2)\,.
\label{eq:Wronskian}
\end{equation}  
In solving the inhomogeneous equation (\ref{eq:ginhomog}) for $\psi$,
we really should impose appropriate side conditions.  Here, the side
conditions we use are not related to boundary conditions in $x$ as
much as analyticity conditions in $\lambda$.  It is easy to check that
for each $x_{0,U}$, the function $\psi_{U}(x,t,\lambda)$ defined by
\begin{equation}
\begin{array}{l}
\displaystyle \psi_U(x,t,\lambda)\,\,=\,\,-\int_{x_{0,U}}^x\frac{g_-(z,t,\lambda)
\exp(-i(\lambda z +\lambda^3 t))\phi(z)}
{\displaystyle \lambda^{1-2N}\prod_{n=1}^N(\lambda^2+4\eta_n^2)}\,dz 
\cdot g_+(x,t,\lambda) \\\\
\displaystyle\hspace{0.8 in}+\,\,
\int_{-\infty}^x \frac{g_+(z,t,\lambda)
\exp(-i(\lambda z +\lambda^3 t))\phi(z)}
{\displaystyle \lambda^{1-2N}\prod_{n=1}^N(\lambda^2 + 4\eta_n^2)}\,dz
\cdot g_-(x,t,\lambda)\,,
\end{array}
\label{eq:psiU}
\end{equation}
is a solution analytic in $\lambda$ for $\Im(\lambda)>0$ and $|\lambda |$
sufficiently large.  Similarly, $\psi_L(x,t,\lambda)$ defined for each
$x_{0,L}$ by
\begin{equation}
\begin{array}{l}
\displaystyle \psi_L(x,t,\lambda)\,\,=\,\,-\int_{x_{0,L}}^x\frac{g_-(z,t,\lambda)
\exp(-i(\lambda z +\lambda^3 t))\phi(z)}
{\displaystyle \lambda^{1-2N}\prod_{n=1}^N(\lambda^2+4\eta_n^2)}\,dz 
\cdot g_+(x,t,\lambda) \\\\
\displaystyle \hspace{0.8 in}-\,\,
\int_x^\infty\frac{g_+(z,t,\lambda)
\exp(-i(\lambda z +\lambda^3 t))\phi(z)}
{\displaystyle \lambda^{1-2N}\prod_{n=1}^N(\lambda^2 + 4\eta_n^2)}\,dz
\cdot g_-(x,t,\lambda)\,,
\end{array}
\label{eq:psiL}
\end{equation}
is a solution analytic for $\Im(\lambda)<0$ and $|\lambda |$
sufficiently large.  The qualification of $|\lambda |$ being
sufficiently large is necessary because the expressions have poles at
the soliton eigenvalues in the respective half-planes where the two
functions $g_\pm(x,t,\lambda)$ become proportional.  However, these
are the only finite singularities and both solutions
$\psi_U(x,t,\lambda)$ and $\psi_L(x,t,\lambda)$ are meromorphic in the
whole of their respective open half-planes.

The arbitrariness of the parameters $x_{0,U}$ and $x_{0,L}$ would seem
to be a problem; but it will turn out that these terms contribute
nothing to the Dunford-Taylor integral that we will prove gives the
required completeness relation.

\subsection{Integrating the resolvent.}
Here we show that the guess we made is indeed correct.
\begin{theorem}
Let $\phi(x)$ be an absolutely continuous function in
$L^1(\mathbb{R})$.  Let $x_{0,U}$ and $x_{0,L}$ be constants and let
$t\in\mathbb{R}$ be fixed.  Then,
\begin{equation}
\phi(x)=-\frac{1}{2\pi i}
\lim_{R\rightarrow\infty}
\left[\int_{C_U}\partial_x\psi_U(x,t,\lambda)\,d\lambda
+ \int_{C_L}\partial_x\psi_L(x,t,\lambda)\,d\lambda\right]\,,
\label{eq:loopexpansion}
\end{equation}
where $C_U$ is the positively oriented half-circle from $R$ to $-R$ in
the upper half-plane, and $C_L$ is the positively oriented half-circle
from $-R$ to $R$ in the lower half-plane.
\label{thm:loop}
\end{theorem}

\noindent{\bf Proof:} First, we show that the terms depending on the
arbitrary parameters $x_{0,U}$ and $x_{0,L}$ converge to zero as
$R\rightarrow\infty$.  This will justify calling the function
$\psi_{U}$ or $\psi_L$ a ``resolvent'' even though the inverse is not
unique.  Consider the integral
\begin{equation}
\begin{array}{rcl}
J_U &:= &\displaystyle
\int_{C_U}\partial_x\left[-\int_{x_{0,U}}^x\frac{g_-(z,t,\lambda)
\exp(-i(\lambda z + \lambda^3 t))\phi(z)}{\displaystyle
\lambda^{1-2N}\prod_{n=1}^N (\lambda^2 + 4\eta_n^2)}\,dz\cdot g_+(x,t,\lambda)\right]\,d\lambda\\\\
&=&\displaystyle
-\int_{C_U}\frac{\displaystyle \lambda^{2N}h_+(x,t,\lambda)}{\displaystyle
\lambda\prod_{n=1}^N(\lambda^2 + 4\eta_n^2)} 
\int_{x_{0,U}}^x g_-(z,t,\lambda)\exp(-i(\lambda z +\lambda^3 t))\phi(z)\,dz\,
d\lambda\,,
\end{array}
\label{eq:JU}
\end{equation}
where we have used the relation (\ref{eq:simplify}).  Recall that
\begin{equation}
\begin{array}{rcl}
h_+(x,t,\lambda)&=&\displaystyle
\sum_{n=1}^{N}\lambda^{-n}\partial_xf_{N-n}(x,t)\,,\\\\
\displaystyle
g_-(z,t,\lambda)\exp(-i(\lambda z+\lambda^3 t))&=&\displaystyle
1+\sum_{n=1}^N(-\lambda)^{-n}
f_{N-n}(z,t)\,.
\end{array}
\end{equation}
Therefore, for all $\lambda$ with $|\lambda|=R>1$,
\begin{equation}
|h_+(x,t,\lambda)|\le \frac{1}{R}\sum_{n=1}^N|\partial_x f_{N-n}(x,t)|\,,
\end{equation}
and 
\begin{equation}
\sup_{z\in\mathbb{R}}|g_-(z,t,\lambda)\exp(-i(\lambda z +\lambda^3 t))|\le
1+\sup_{z\in\mathbb{R}}\sum_{n=1}^N|f_{N-n}(z,t)|\,.
\end{equation}
This latter relation assumes the uniform boundedness of the functions
$f_k(z,t)$ in $z$.  Finally, it is clear that for $|\lambda|=R>\sup_n
2\eta_n$,
\begin{equation}
\left|\lambda^{1-2N}\prod_{n=1}^N(\lambda^2+4\eta_n^2)\right|\ge 
R\prod_{n=1}^N \left(1-\frac{4\eta_n^2}{R^2}\right),.
\end{equation}
It follows that for all $\lambda$ with $|\lambda|=R$ sufficiently large,
\begin{equation}
|J_U|\le \frac{K(x,t)}{R}\|\phi\|_1\,,
\label{eq:simplebound}
\end{equation}
where
\begin{equation}
K(x,t)=\pi \prod_{n=1}^N
\left(1-\frac{4\eta_n^2}{R^2}\right)^{-1}\cdot
\left(\sum_{n=1}^N|\partial_xf_{N-n}(x,t)|\right)\cdot
\left(1+\sup_{z\in\mathbb{R}}\sum_{n=1}^N|f_{N-n}(z,t)|\right)
\,.
\end{equation}
The bound (\ref{eq:simplebound}) clearly vanishes as
$R\rightarrow\infty$.  A nearly identical argument shows that the
integral
\begin{equation}
\begin{array}{rcl}
J_L&:= &\displaystyle \int_{C_L}\partial_x\left[-\int_{x_{0,L}}^x
\frac{g_-(z,t,\lambda)\exp(-i(\lambda z+\lambda^3 t))\phi(z)}{
\displaystyle \lambda^{1-2N}\prod_{n=1}^N(\lambda^2+4\eta_n^2)}\,dz\cdot
g_+(x,t,\lambda)\right]\,d\lambda
\end{array}
\label{eq:JL}
\end{equation}
satisfies the same bound (\ref{eq:simplebound}) as $J_U$.

Now we consider integrating the second terms of
$\partial_x\psi_U(x,t,\lambda)$ and $\partial_x\psi_L(x,t,\lambda)$
respectively.  For brevity, define
\begin{equation}
Y(x,z,t,\lambda):= \frac{g_+(z,t,\lambda)\exp(-i(\lambda z
+\lambda^3 t)) h_-(x,t,\lambda)} {\displaystyle
\lambda^{1-2N}\prod_{n=1}^N(\lambda^2 +4\eta_n^2)}\,.
\label{eq:Ydef}
\end{equation}
Note that this can be written as
\begin{equation}
\begin{array}{l}
\displaystyle
Y(x,z,t,\lambda)=\exp(i\lambda(x-z))\frac{\displaystyle
\left(1+\sum_{n=1}^N\frac{f_{N-n}(z,t)}{\lambda^n}\right)}
{\displaystyle \lambda\prod_{n=1}^N\left(1+\frac{4\eta_n^2}{\lambda^2}\right)}
\\\\
\displaystyle\hspace{0.3 in}\times\,\,
\left(i\lambda\left(1+\sum_{n=1}^N\frac{f_{N-n}(x,t)}
{(-\lambda)^n}\right)+
\sum_{n=1}^N \frac{\partial_xf_{N-n}(x,t)}{(-\lambda)^n}\right)

\,,
\end{array}
\label{eq:Yexpand}
\end{equation}
and therefore,
\begin{equation}
Y(x,z,t,\lambda)=i\exp(i\lambda(x-z))\left(1+\Delta(x,z,t,\lambda))\right)\,,
\label{eq:Yapprox}
\end{equation}
where $\Delta(x,z,t,\lambda)=O(\lambda^{-1})$ uniformly in $x$ and $z$
for fixed $t$.  It also follows from additional cancellation that for
$z=x$, $\Delta(x,x,t,\lambda)=O(\lambda^{-2})$ uniformly in $x$.
Finally, derivatives of $\Delta$ are controlled as well:
$\partial_z\Delta(x,z,t,\lambda)=O(\lambda^{-1})$ uniformly.  The
integral we need to compute for the contribution of
$\partial_x\psi_U(x,t,\lambda)$ is
\begin{equation}
\begin{array}{l}
\displaystyle \int_{C_R}\int_{-\infty}^x
Y(x,z,t,\lambda)\phi(z)\,dz\,d\lambda =
i\int_{C_R}\int_{-\infty}^x \exp(i\lambda(x-z))\phi(z)\,dz\, d\lambda\\\\
\displaystyle \hspace{0.4 in}+\,\,
i\int_{C_R}\int_{-\infty}^x
\Delta(x,z,t,\lambda)\exp(i\lambda(x-z))\phi(z)\,dz
\,d\lambda\,.
\end{array}
\label{eq:nextintegral}
\end{equation}
Note that since the integrand is analytic in the upper half-plane, the
first term can be written as:
\begin{equation}
i\int_{C_R}\int_{-\infty}^x\exp(i\lambda(x-z))\phi(z)\,dz\,d\lambda =
-i\int_{-R}^R\int_{-\infty}^x\exp(i\lambda(x-z))\phi(z)\,dz\,d\lambda\,.
\label{eq:FourierHalf}
\end{equation}
In order to control the error term, it is necessary to integrate by parts once:
\begin{equation}
\begin{array}{l}
\displaystyle i\int_{C_R}\int_{-\infty}^x\Delta(x,z,t,\lambda)
\exp(i\lambda(x-z))\phi(z)\,dz\,d\lambda =
-\phi(x)\int_{C_R}\frac{\Delta(x,x,t,\lambda)}{\lambda}\,d\lambda \\\\
\displaystyle\hspace{0.4 in}+\,\,\int_{C_R}\int_{-\infty}^x
\frac{\exp(i\lambda(x-z))}{\lambda}\partial_z(\Delta(x,z,t,\lambda)\phi(z))\,dz
\,d\lambda\,.
\end{array}
\label{eq:intbyparts}
\end{equation}
The boundary term at $z=-\infty$ vanishes because
$\Delta(x,z,t,\lambda)$ is bounded there, $\phi$ is continuous and
integrable, and $\exp(i\lambda(x-z))$ is exponentially small for
$\Im(\lambda)>0$.  Since the exponential is bounded in magnitude by
unity for $\Im(\lambda)>0$, and the contour is of length $\pi R$, the
above estimates of $\Delta$ imply that there exist $K_0(x,t)$,
$K_1(x,t)$ and $K_2(x,t)$ all positive, such that
\begin{equation}
\begin{array}{l}
\displaystyle \left|\int_{C_R}\int_{-\infty}^x\Delta(x,z,t,\lambda)
\exp(i\lambda(x-z))\phi(z)\,dz\,d\lambda\right| =\\\\
\displaystyle\hspace{0.4 in}\frac{K_0(x,t)}{R^2}|\phi(x)|+
\frac{K_1(x,t)}{R}\|\phi\|_1 +\frac{K_2(x,t)}{R}\|\phi'\|_1\,.
\end{array}
\label{eq:errorbound}
\end{equation}
This proves that
\begin{equation}
\lim_{R\rightarrow\infty}\int_{C_R}\int_{-\infty}^xY(x,z,t,\lambda)\phi(z)\,dz
\,d\lambda = 
-i\lim_{R\rightarrow\infty}\int_{-R}^R\int_{-\infty}^x
\exp(i\lambda(x-z))\phi(z)\,dz\,d\lambda\,.
\label{eq:limitforpsiU}
\end{equation}
Similar arguments applied to the contribution of
$\partial_x\psi_L(x,t,\lambda)$ show that
\begin{equation}
-\lim_{R\rightarrow\infty}\int_{C_L}\int_x^\infty Y(x,z,t,\lambda)\phi(z)\,dz
\,d\lambda = 
-i\lim_{R\rightarrow\infty}\int_{-R}^R\int_x^\infty
\exp(i\lambda(x-z))\phi(z)\,dz\,d\lambda\,,
\label{eq:limitforpsiL}
\end{equation}
and therefore
\begin{equation}
\begin{array}{l}
\displaystyle
-\frac{1}{2\pi i}
\lim_{R\rightarrow\infty}\left[\int_{C_R}\partial_x
\psi_U(x,t,\lambda)\,d\lambda+
\int_{C_L}\partial_x\psi_L(x,t,\lambda)\,d\lambda\right]=\\\\
\hspace{0.4 in}\displaystyle\frac{1}{2\pi}\lim_{R\rightarrow\infty}
\int_{-R}^R\int_{-\infty}^\infty
\exp(i\lambda(x-z))\phi(z)\,dz\,d\lambda=\phi(x)\,,
\end{array}
\label{eq:putittogether}
\end{equation}
with the last equality following from Fourier inversion.  This
establishes (\ref{eq:loopexpansion}) and the theorem.  \pfend

As it stands, the completeness relation given in
Theorem~\ref{thm:loop} is not really an expansion of $\phi(x)$ in
terms of the functions $h_-(x,t,\lambda)$, because the expansion
coefficients themselves depend on $x$.  This is easily remedied by
casting the right-hand side of the completeness relation into a more
useful form.  We do this now.
\begin{theorem}
Let $\phi(x)$ be an absolutely continuous function in
$L^1(\mathbb{R})$. Let $t\in\mathbb{R}$ be fixed, and choose any
$w\in\mathbb{R}\cup\{-\infty,+\infty\}$.  Define the mode function
\begin{equation}
H(x,t,\lambda):= \lambda^Nh_-(x,t,\lambda)\,,
\label{eq:mode}
\end{equation}
which is an entire function of $\lambda$, and the amplitudes
\begin{equation}
\begin{array}{rcl}
b^+(t,\lambda)&:= &\displaystyle
\int_w^\infty 
\frac{\lambda^N g_+(z,t,\lambda)\exp(-i(\lambda z +\lambda^3 t))
}{\displaystyle\lambda\prod_{n=1}^N(\lambda^2 + 4\eta_n^2)}\phi(z)\,dz\,,\\\\
b^-(t,\lambda)&:= &\displaystyle
\int_{-\infty}^w 
\frac{\lambda^N g_+(z,t,\lambda)\exp(-i(\lambda z +\lambda^3 t))
}{\displaystyle\lambda\prod_{n=1}^N(\lambda^2 + 4\eta_n^2)}\phi(z)\,dz\,,
\end{array}
\label{eq:amplitudes}
\end{equation}
and set $b(t,\lambda):=b^+(t,\lambda)+b^-(t,\lambda)$.  The amplitudes have
simple poles at $\lambda=0$ and $\lambda=\pm 2i\eta_n$ for
$n=1,\dots,N$.  Finally, set
\begin{equation}
b_0(t):=\frac{1}{2}\res{\lambda=0}(b^+(t,\lambda)-b^-(t,\lambda))\,,
\hspace{0.2 in}\mbox{ and }\hspace{0.2 in}
b^\pm_n(t):=\mp\res{\lambda=\pm 2i\eta_n}b^\mp(t,\lambda)\,.
\label{eq:residues}
\end{equation}
Then we have the expansion
\begin{equation}
\begin{array}{l}
\displaystyle
\phi(x)=\lim_{R\rightarrow\infty}\frac{1}{2\pi i}{\rm P.\,V.\,}
\int_{-R}^R b(t,\lambda)H(x,t,\lambda)\,d\lambda \\\\
\displaystyle\hspace{0.5 in}+\,\, b_0(t)
H(x,t,0) 
+ \sum_{n=1}^N \left[b^-_n(t) 
H(x,t,-2i\eta_n) + b^+_n(t)
H(x,t,2i\eta_n)\right]\,.
\end{array}
\label{eq:expansion}
\end{equation}
\label{thm:completeness}
\end{theorem}

\noindent{\bf Remark:} Since $w$ is now fixed and not a function of
$x$, this expansion (\ref{eq:expansion}) is a true completeness
relation, expressing an arbitrary given function $\phi(x)$ as a sum of
known functions $H(x,t,\lambda)$.  From the exact formulas
(\ref{eq:mode}), (\ref{eq:particularsolutions}), and
(\ref{eq:formoff}), it is clear that the part of the expansion
(\ref{eq:expansion}) represented by the singular integral is
Fourier-like, with the corresponding components of the solution,
$H(x,t,\lambda)$ for $\lambda\in\mathbb{R}$ being bounded oscillatory
functions tending to complex exponentials for large $x$.  On the other
hand, the discrete contributions to the solution represent bound
states.  The $2N+1$ bound state terms in (\ref{eq:expansion}) are not
linearly independent.  From the fact that at the eigenvalues $\pm
2i\eta_n$ the functions $g_-(x,t,\lambda)$ are all linear combinations
of the same $N$ functions $f_0(x,t),\dots,f_{N-1}(x,t)$, it is clear
that only $N$ of the bound states are linearly independent.  These
facts are easiest to see when one takes $w$ to $\infty$ or $-\infty$.
Then, half of the contributions from the eigenvalues disappear, and it
only remains to express the bound state at zero, $H(x,t,0)$, in terms
of $H(x,t,\pm 2i\eta_n)$.  This can be done directly.  From the exact
formulas (\ref{eq:mode}), (\ref{eq:particularsolutions}) and
(\ref{eq:formoff}), we see that
\begin{equation}
H(x,t,0)=(-1)^N\partial_xf_0(x,t)\,,
\end{equation}
and making use of the relations (\ref{eq:KMrelations}) satisfied by
$f_+$ at the eigenvalues,
\begin{equation}
H(x,t,2i\eta_n)=(-1)^{n+1}\exp(-2\eta_n\alpha_n)\sum_{p=0}^{N-1} (2i\eta_n)^p\partial_xf_p(x,t)\,.
\end{equation}
Expressing $H(x,t,0)$ in terms of $H(x,t,2i\eta_n)$ is therefore a
polynomial interpolation problem.  Introduce the polynomial 
\begin{equation}
P(\lambda)=\sum_{p=0}^{N-1}\partial_xf_p(x,t)\lambda^p\,.
\end{equation}
Given isolated values of this polynomial:
\begin{equation}
P(2i\eta_n)=(-1)^{n+1}\exp(2\eta_n\alpha_n)H(x,t,2i\eta_n)\,,\hspace{0.2 in}
\mbox{ for }\hspace{0.2 in}n=1,\dots,N\,,
\end{equation}
we are to find $P(0)$ and thus
$H(x,t,0)=(-1)^N\partial_xf_0(x,t)=(-1)^NP(0)$.  Expressing
$P(\lambda)$ explicitly in terms of Lagrange polynomials gives
\begin{equation}
P(\lambda)=\sum_{n=1}^N(-1)^{n+1}\exp(2\eta_n\alpha_n)H(x,t,2i\eta_n)\prod_{k\neq n}\frac{\lambda-2i\eta_k}{2i\eta_n-2i\eta_k}\,,
\end{equation}
and therefore
\begin{equation}
H(x,t,0)=\sum_{n=1}^N \left[(-1)^n\exp(2\eta_n\alpha_n)\prod_{k\neq
n}\frac{\eta_k}{\eta_n-\eta_k}\right]H(x,t,2i\eta_n)\,.
\end{equation}
There is of course a similar expression for $H(x,t,0)$ in terms of
$H(x,t,-2i\eta_n)$.  \pfend

\noindent{\bf Remark:} An important distinction between the
completeness relation stated in Theorem~\ref{thm:completeness} and
that found by Sachs \cite{S83} for derivatives of squared Schr\"odinger
eigenfunctions is in the nature of the singularity at $\lambda=0$.
Sachs shows that in the expansion of $\phi$ in terms of derivatives of
squared eigenfunctions, there is an apparent singularity at
$\lambda=0$ that it is in fact removable.  On the other hand, the
integral in Theorem~\ref{thm:completeness} is essentially singular and
the residue contribution of $b_0(t)$ is nonzero.  \pfend

\noindent{\bf Proof of Theorem~\ref{thm:completeness}:} We first
establish that in the formulas (\ref{eq:psiU}) for
$\partial_x\psi_U(x,t,\lambda)$ and (\ref{eq:psiL}) for
$\partial_x\psi_L(x,t,\lambda)$ we may replace $x$ in the limits of
integration by any other value without changing the result of the
theorem.  That is, we will now show that the integral
\begin{equation}
\int_{C_U}\int_{-\infty}^w Y(x,z,t,\lambda)\phi(z)\,dz\,d\lambda -
\int_{C_L}\int_w^\infty Y(x,z,t,\lambda)\phi(z)\,dz\,d\lambda\,,
\end{equation}
which we have already seen converges as $R$ tends to infinity to
$-2\pi i \phi(x)$ in the case that $w=x$, is in fact independent of
$w$.  Holding $R$ fixed and differentiating with respect to $w$, we must 
show that for sufficiently large $R$, 
\begin{equation}
\phi(w)\oint_{|\lambda|=R}Y(x,w,t,\lambda)\,d\lambda \equiv 0\,,
\end{equation}
identically in $x$, $w$, and $t$.  Being as the integrand is
meromorphic in the finite $\lambda$ plane, we can evaluate the
integral by residues.  There are simple poles at $\lambda=0$ and
$\lambda=\pm 2i\eta_n$ for $n=1, \dots,N$.  Using the linear relations
(\ref{eq:KMrelations}) satisfied by $f_\pm$ at the eigenvalues
$\lambda=2i\eta_n$, we find
\begin{equation}
\begin{array}{rcl}
\displaystyle\res{\lambda=2i\eta_k}Y(x,w,t,\lambda)&=&\displaystyle
\sum_{p=0}^{N-1}
\frac{(2i\eta_k)^{N+p-1}}
{D_k}\partial_x f_p(x,t)\\\\
&&\displaystyle\hspace{0.4 in}+\,\,
\sum_{p,q=0}^{N-1} (-1)^{N-q}\frac{(2i\eta_k)^{p+q-1}}
{D_k}f_q(w,t)\partial_xf_p(x,t)\,,\\\\
\displaystyle
\res{\lambda=-2i\eta_k}Y(x,w,t,\lambda)&=&\displaystyle
\sum_{p=0}^{N-1}
(-1)^{N-p}\frac{(2i\eta_k)^{N+p-1}}{D_k}
\partial_xf_p(x,t)\\\\
&&\displaystyle\hspace{0.4 in}+\,\,
\sum_{p,q=0}^{N-1}(-1)^{N-p} \frac{(2i\eta_k)^{p+q-1}}{D_k}
f_q(w,t)\partial_xf_p(x,t)\,,
\end{array}
\end{equation}
where 
\begin{equation}
D_k:= \prod_{n\neq k}(2i\eta_k-2i\eta_n)\prod_{n=1}^N(2i\eta_k+2i\eta_n)\,.
\end{equation}
Similarly, for the residue at zero,
\begin{equation}
\res{\lambda=0}Y(x,w,t,\lambda)=(-1)^N\frac{f_0(w,t)\partial_xf_0(x,t)}
{\displaystyle\prod_{n=1}^N4\eta_n^2}\,.
\end{equation}
Adding all the residues and collecting coefficients of the terms
$\partial_xf_p(x,t)$ and $f_q(w,t)\partial_xf_p(x,t)$, we find that the sum
of the residues will be zero if:
\begin{equation}
I_p:=\sum_{k=1}^N\frac{(2i\eta_k)^p}{D_k}
 = 0
\end{equation}
for all odd $p=1,3,5,\dots,2N-3$, and if:
\begin{equation}
\frac{1}{\displaystyle\prod_{n=1}^N4\eta_n^2} + 2\sum_{k=1}^N
\frac{1}{2i\eta_kD_k} = 0\,.
\end{equation}
These expressions are themselves sums of residues of {\em
meromorphic} differentials.  Thus, by inspection, one finds that for
$p=1,3,5,\dots,2N-3$,
\begin{equation}
I_p=\frac{1}{2\pi i}\oint_C \frac{\lambda^p\,d\lambda}{\displaystyle
\prod_{n=1}^N(\lambda^2+4\eta_n^2)}\,,
\end{equation}
where $C$ is any simple counter-clockwise oriented contour that encircles the
points $\lambda=2i\eta_n$ for $n=1,\dots,N$ (but without enclosing the
conjugate eigenvalues or $\lambda=0$).  With $p$ bounded by $2N-3$, the path
of integration can be blown out to infinity in the upper half-plane, and then
brought down to the real axis so that
\begin{equation}
I_p=\frac{1}{2\pi i}\int_{-\infty}^\infty \frac{\lambda^p\,d\lambda}
{\displaystyle \prod_{n=1}^N(\lambda^2+4\eta_n^2)} = 0\,,
\end{equation}
with the last equality following from the oddness of the integrand for
odd $p$.  Finally, consider the integral $I_{-1}$ defined by
\begin{equation}
I_{-1}:=\frac{1}{2\pi i}\oint_C\frac{d\lambda}{\displaystyle\lambda
\prod_{n=1}^N(\lambda^2+4\eta_n^2)}\,.
\end{equation}
Evaluating the residues inside $C$, we find
\begin{equation}
I_{-1}=\sum_{k=1}^N\frac{1}{\displaystyle 2i\eta_k\prod_{n\neq k}(2i\eta_k-2i\eta_n)\prod_{n=1}^N(2i\eta_k+2i\eta_n)}\,.
\end{equation}
On the other hand, we can again blow the contour $C$ out to infinity
in the upper half-plane and bring it down to the real axis.  This
time, there is a singularity at $\lambda=0$, so the Plemelj formula
must be used.  We find
\begin{equation}
I_{-1}=-\frac{1}{2}\cdot\frac{1}{\displaystyle\prod_{n=1}^N4\eta_n^2} +
\frac{1}{2\pi i}{\rm P.\,V.\,}\int_{-\infty}^\infty \frac{d\lambda}{
\displaystyle \lambda\prod_{n=1}^N(\lambda^2 +4\eta_n^2)}\,.
\end{equation}
Once again, by oddness, the principal value integral vanishes
identically, and then combining this result with the previous
expression we obtain the required vanishing.  

This shows that for any $w$,
\begin{equation}
\phi(x)=-\frac{1}{2\pi i}\lim_{R\rightarrow\infty}\left[
\int_{C_U}\int_{-\infty}^w Y(x,z,t,\lambda)\phi(z)\,dz\,d\lambda -
\int_{C_L}\int_w^\infty Y(x,z,t,\lambda)\phi(z)\,dz\,d\lambda\right]\,.
\label{eq:expansionw}
\end{equation}
Establishing (\ref{eq:expansion}) and therefore the Theorem now
amounts to using the residue theorem once again to deform the
integration paths $C_U$ and $C_L$ in (\ref{eq:expansionw}) to the real
axis.  One finds discrete contributions at the poles $\lambda=\pm
2i\eta_n$ and then applying the Plemelj formula to contract the
contour to the real axis in the neighborhood of $\lambda=0$ gives a
discrete contribution proportional to $H(x,t,0)$ and the principal
value regularization of the singular integral over the continuous
spectrum. \pfend

\section{Solution of the Initial Value Problem for $\kappa=1/2$}
\label{sec:general}
It is easy to see that when $A(x,t)$ is an $N$-soliton solution of KdV
(\ref{eq:kdv}), one can use the completeness relation to solve the initial
value problem:
\begin{equation}
\partial_tB+\partial_x\left[\frac{1}{2}AB+\partial_x^2 B\right]=0\,,
\hspace{0.3 in}B(x,0)=\phi(x)\,.
\label{eq:linearagain}
\end{equation}
Setting $t=0$, and picking a convenient value of $w$, say,
$w=+\infty$, one computes the amplitudes (\ref{eq:amplitudes}) and
discrete coefficients (\ref{eq:residues}).  Then, because the function
$H(x,t,\lambda)$ satisfies (\ref{eq:linear}) for $\kappa=1/2$ and for
each complex $\lambda$, the expression
\begin{equation}
\begin{array}{l}
\displaystyle
B(x,t):=\lim_{R\rightarrow\infty}\frac{1}{2\pi i}{\rm P.\,V.\,}
\int_{-R}^R b(0,\lambda)H(x,t,\lambda)\,d\lambda \\\\
\displaystyle\hspace{0.5 in}+\,\, b_0(0)
H(x,t,0) 
+ \sum_{n=1}^N \left[b^-_n(0) 
H(x,t,-2i\eta_n) + b^+_n(0)
H(x,t,2i\eta_n)\right]\,,
\end{array}
\label{eq:IVPsolution}
\end{equation}
provides the solution of the initial value problem
(\ref{eq:linearagain}), generally in the sense of distributions.  That
is, $B(x,0)=\phi(x)$ by Theorem~\ref{thm:completeness}, and for each
test function $\varphi(x,t)$ that is differentiable in $t$ and three
times differentiable in $x$ and has compact support in
$(x,t)\in\mathbb{R}\times\mathbb{R}_+$, one shows by exchanging the
order of integration that
\begin{equation}
\int_0^\infty\int_{-\infty}^\infty\left[\partial_t\varphi(x,t) +\frac{1}{2}A(x,t)
\partial_x\varphi(x,t) +
\partial_x^3\varphi(x,t)\right]B(x,t)\,dx\,dt=0\,.
\end{equation}
The solution will be classical in as much as it is possible to
differentiate with respect to $x$ and $t$ under the integral sign in
the solution formula (\ref{eq:IVPsolution}).  This requires additional
smoothness and decay assumptions on the initial data $\phi(x)$ that we
do not consider here.

\section{Scattering of Bound States for $\kappa=1/2$}
\label{sec:scattering}
Of particular interest in applications is the $N$-dimensional (recall
that $A(x,t)$ is an $N$-soliton solution of KdV) subspace of solutions
of (\ref{eq:linear}) for $\kappa=1/2$ consisting of bound states.
This subspace represents linear waves that are trapped by the solitons
of the potential $A(x,t)$.  For large $|t|$, these bound state
solutions are all confined to the trajectories of the solitons.
Therefore, it follows that each bound state $B(x,t)$ has two
asymptotic representations:
\begin{equation}
B(x,t)\sim \sum_{n=1}^N \beta_n^\pm A^\pm_n(x,t)\,,\hspace{0.4 in}
t\rightarrow \pm\infty\,,
\end{equation}
for some constants $\beta_n^\pm$ depending on $B(x,t)$, where
$A_n^\pm(x,t)$ are defined by (\ref{eq:Andef}).  Since there are
exactly $N$ linearly independent bound states, it follows that the
constants $\beta^+_n$ are completely determined from the constants
$\beta_n^-$.  In particular, there exists an invertible $N\times N$
matrix $\bf T$ with entries depending only on the data specifying the
$N$-soliton solution $A(x,t)$, such that
\begin{equation}
\beta_j^+=\sum_{k=1}^N T_{jk}\beta_k^-\,.
\end{equation}
The matrix $\bf T$ is called the {\em bound state scattering matrix}.
In this section, we compute the scattering matrix explicitly, and show
that its elements only depend on the soliton eigenvalues
$\eta_1,\dots,\eta_N$.

If $A(x,t)$ is an $N$-soliton solution of KdV (\ref{eq:kdv}), then
a family of solutions to (\ref{eq:linear}) for $\kappa=1/2$, 
parametrized by complex $\lambda$, is given by
\begin{equation}
h_+(x,t,\lambda)=\frac{A(x,t)}{6i\lambda} + \sum_{n=0}^{N-2}\lambda^{n-N}
\partial_xf_n(x,t)\,.
\label{eq:boundstate}
\end{equation}
We want to analyze these solutions in the limit of large $|t|$, in a
frame of reference traveling with constant velocity $c$.

The first step is to see how the coefficients $f_n(x,t)$ behave for
large $|t|$.  Let $\chi=x-ct$ be fixed as $\tau=t$ goes to either
$+\infty$ or $-\infty$.  Begin by taking $\eta_m^2<4c<\eta_{m-1}^2$ to
see how the coefficients behave in between the solitons.  In the limit
$\tau\rightarrow+\infty$, the equations (\ref{eq:KMrelations}) imply
that
\begin{equation}
\begin{array}{rccccll}
1&+&\displaystyle \sum_{k=0}^{N-1}(-2i\eta_n)^{k-N}f_k&
\rightarrow &0\,,&
n=1,\dots,m-1\,,\\
\\
1&+&\displaystyle \sum_{k=0}^{N-1}(2i\eta_n)^{k-N}f_k
&\rightarrow &0\,,&
n=m,\dots,N\,.
\end{array}
\end{equation}
This is an invertible Vandermonde system for the coefficients $f_k$,
so that as $\tau\rightarrow +\infty$, the $f_k$ all become constants,
independent of $\chi$ and $\tau$.  Thus, $\partial_xf_k(x,t)$ vanishes
between the solitons for all $k$.  The analogous result holds as
$\tau\rightarrow-\infty$.  This shows that the solutions of
(\ref{eq:linear}) for $\kappa=1/2$ described by the formula
(\ref{eq:boundstate}) are asymptotically confined to the individual
frames of reference of the moving solitons in the potential field
$A(x,t)$.

Now set $c=4\eta_m^2$ to go into the moving frame of reference of
one of the solitons.  Taking the limit $\tau\rightarrow+\infty$ yields
\begin{equation}
\begin{array}{rccccll}
1&+&\displaystyle \sum_{k=0}^{N-1}(-2i\eta_n)^{k-N}f_k&\rightarrow &0\,,&
n=1,\dots,m-1\,,\\
\\
1&+&\displaystyle \sum_{k=0}^{N-1}(2i\eta_n)^{k-N}f_k&\rightarrow &0\,,&
n=m+1,\dots,N\,.
\end{array}
\end{equation}
This is a system of $N-1$ equations in $N$ unknowns, so it can be used
to asymptotically eliminate $\partial_\chi f_0$ through $\partial_\chi
f_{N-2}$ in favor of $\partial_\chi f_{N-1}$, which we know is
proportional to the $N$-soliton solution of KdV, $A(x,t)$.  Thus, as
$\tau\rightarrow +\infty$, with $c=4\eta_m^2$,
\begin{equation}
\partial_\chi f_k = Q_{mk}\partial_\chi f_{N-1} = 
\frac{1}{6i}Q_{mk}A\,,
\end{equation}
for $k=0,\dots,N-2$ where the numbers $Q_{mk}$ are the unique solution
of the inhomogeneous system of linear algebraic equations
\begin{equation}
\begin{array}{rccccll}
(-2i\eta_n)^{-1}&+&\displaystyle \sum_{k=0}^{N-2}(-2i\eta_n)^{k-N}Q_{mk}&=&
 0\,,&
n=1,\dots,m-1\,,\\
\\
(2i\eta_n)^{-1}&+&\displaystyle \sum_{k=0}^{N-2}(2i\eta_n)^{k-N}Q_{mk}&=& 0\,,&
n=m+1,\dots,N\,.
\end{array}
\end{equation}
One can similarly show that as $\tau\rightarrow -\infty$, with
$c=4\eta_m^2$,
\begin{equation}
\partial_\chi f_k=Q_{mk}^*\partial_\chi 
f_{N-1} = \frac{1}{6i}Q_{mk}^*A\,,
\end{equation}
for $k=0,\dots,N-2$, where the star denotes complex conjugation.

Now consider particular solutions $B_j(x,t)$ of (\ref{eq:linear}) for
$\kappa=1/2$ obtained as linear combinations of $N$ others expressed
by the formula (\ref{eq:boundstate}) evaluated on the $N$ soliton
eigenvalues.  The formula for $B_j(x,t)$ is
\begin{equation}
B_j(x,t)= \sum_{k=1}^N F_{jk}h_+(x,t,2i\eta_k)
= \sum_{k=1}^N
F_{jk}\left[-\frac{A(x,t)}{12\eta_k} +
\sum_{n=0}^{N-2}(2i\eta_k)^{n-N}\partial_x f_n(x,t) \right]\,,
\end{equation}
where ${\bf F}=\{ F_{jk}\}$ is a matrix of arbitrary constants.  From
the asymptotics of $f_n(x,t)$, we have as $\tau\rightarrow-\infty$
with $c=4\eta_m^2$,
\begin{equation}
B_j\rightarrow A\sum_{k=1}^N F_{jk}G^-_{km}\,,\hspace{0.2 in}
{\rm where}\hspace{0.2 in}
G^-_{km}:= -\frac{1}{12\eta_k}+\frac{1}{6i}\sum_{n=0}^{N-2}(2i\eta_k)^{n-N}
Q_{mn}^*\,.
\end{equation}
So, with the choice that the matrix $\{F_{jk}\}$ is the inverse of the
matrix ${\bf G}^-=\{G^-_{km}\}$, the particular solution $B_j(x,t)$ of
(\ref{eq:linear}) for $\kappa=1/2$ will be completely confined as
$t\rightarrow -\infty$ to the frame of reference moving with speed
$c=4\eta_j^2$, where it will be locally indistinguishable from the
solution $A(x,t)$ of KdV.  Let us now determine how $B_j(x,t)$ will
behave in the various soliton frames as $t\rightarrow +\infty$.
Passing to the limit of $\tau\rightarrow +\infty$ in the frame
with velocity $c=4\eta_m^2$ gives
\begin{equation}
B_j\rightarrow A\sum_{k=1}^N F_{jk}G^+_{km}\,,\hspace{0.2 in}
{\rm where}\hspace{0.2 in}G^+_{km}:= -\frac{1}{12\eta_k}+\frac{1}{6i}
\sum_{n=0}^{N-2}(2i\eta_k)^{n-N}Q_{mn}\,.
\end{equation}
These asymptotics give us a formula for the bound state scattering matrix:
\begin{equation}
{\bf T}:= \left[({\bf G^-})^{-1}{\bf G^+}\right]^T\,.
\label{eq:amplitudetransfer}
\end{equation}
It is clear that the elements of $\bf T$ depend only on the $N$
soliton eigenvalues $\eta_1,\dots,\eta_N$.  There is no dependence on
the soliton phase variables $\alpha_1,\dots,\alpha_N$.  Therefore, the
asymptotic scattering properties of linear waves in (\ref{eq:linear})
with $\kappa=1/2$ are insensitive to phase shifts among the solitons in
the potential $A(x,t)$.  As a concrete example of the scattering matrix, we 
compute it explicitly for $N=2$ for arbitrary $\eta_1>\eta_2>0$:
\begin{equation}
{\bf T}=\frac{1}{\eta_1^2-\eta_2^2}\left[\begin{array}{cc}
(\eta_1-\eta_2)^2 & 2\eta_2(\eta_1-\eta_2) \\
2\eta_1(\eta_1-\eta_2) & -(\eta_1-\eta_2)^2
\end{array}\right]\,.
\end{equation}
The fact that $T_{22}$ is negative means that it is possible for the
interactions of the solitons in $A(x,t)$ to convert trapped linear waves of
elevation into waves of depression, and vice-versa.

\section{General Values of $\kappa$}
\label{sec:numerics}
We expect that for most values of $\kappa$, the linear waves
satisfying (\ref{eq:linear}) will not be permanently trapped by
solitons present in the potential $A(x,t)$.  This is suggested by
considering the simplest case, namely taking $A(x,t)$ to be the
one-soliton solution of KdV (\ref{eq:kdv}).  The soliton travels with
velocity $c=4\eta^2$, so that $A=-V(\chi)$ with $\chi=x-ct-\alpha$.
Corresponding traveling wave solutions $B(\chi)$ of the linear problem
that propagate with the same velocity and decay as
$\chi\rightarrow\pm\infty$ satisfy
\begin{equation}
-B''(\chi)+\kappa V(\chi) B(\chi) = -cB(\chi)\,.
\label{eq:ev}
\end{equation}
Since $c$ is fixed, we can view this as an eigenvalue equation with
$\kappa$ as the eigenvalue.  We therefore expect that only isolated
values of $\kappa$ will admit nontrivial decaying solutions $B(\chi)$.
We have already seen that $\kappa=1/2$ and $\kappa=1$ are indeed
eigenvalues.  For $\kappa=1/2$ the eigenfunction $B(\chi)$ is an even
function of $\chi$, while for $\kappa=1$ the eigenfunction $B(\chi)$
is odd in $\chi$.  Since eigenfunctions of (\ref{eq:ev}) must be
nondegenerate and therefore have either odd or even parity in $\chi$,
there cannot exist a nontrivial bound state eigenfunction of
(\ref{eq:ev}) for all $\kappa\in[1/2,1]$ since the eigenfunction would
have to change parity from one endpoint to the other.  Therefore, at
least one value of $\kappa\in[1/2,1]$ is not an eigenvalue.  For such
$\kappa$, there is no bound state traveling wave solution of
(\ref{eq:linear}) that is trapped in the soliton trajectory.

We can be more precise about this phenomenon.  The left-hand side of
(\ref{eq:ev}) can also be viewed as a Schr\"odinger operator
$L(\kappa)$ depending on a coupling constant $\kappa$, and the
condition for wave trapping by solitons is simply that
$-c\in\Sigma_{\rm p}(L(\kappa))$, where $\Sigma_{\rm p}$ denotes the
point spectrum.  The number of discrete eigenvalues is a nondecreasing
function of $\kappa>0$, corresponding to the deepening of the
potential well.  There is an infinite unbounded sequence of {\em
cutoff} values $\kappa^{\rm cut}_n$ of $\kappa$ at which the number of
eigenvalues changes by one, and the new eigenvalue is born from the
continuum.  Each eigenvalue, once born, is distinct and is a
decreasing function of $\kappa$.  From these arguments, it follows
that there exists an infinite unbounded sequence of {\em bifurcation}
values $\kappa^{\rm bif}_n$ of $\kappa$ at which one eigenvalue crosses
the level $E=-c$, and a bound state traveling wave solution of
(\ref{eq:linear}) exists.

It is easy to find the bifurcation points because the hyperbolic
secant squared potential is so well understood.  The potential $\kappa
V(\chi)$ is exactly reflectionless for $12\kappa=n(n+1)$ for
$n=1,2,3,\dots$.  The corresponding energy levels are
$E_{n,k}=-k^2\eta^2$, for $k=1,\dots,n$.  Therefore, for $n>1$ in this
sequence, there is always one eigenvalue that is exactly equal to $-c
= -4\eta^2$.  The corresponding eigenstate is always the $n-1$st state
and therefore has $n-2$ zeros.  It follows that the bifurcation points are
$\kappa=\kappa^{\rm bif}_n =(n+1)(n+2)/12$ for $n=1,2,3,\dots$.

The fact that some linear waves may be permanently trapped by isolated
solitons at a bifurcation point $\kappa=\kappa^{\rm bif}_n$ does not
necessarily imply that there will be no losses to radiation when
solitons in the field $A(x,t)$ interact with one another.  Such a
lossless interaction might suggest the ``integrability'' of the linear
equation (\ref{eq:linear}).  We have indeed seen that this is the case
for the first two bifurcation points, $\kappa=\kappa^{\rm bif}_1=1/2$
and $\kappa=\kappa^{\rm bif}_2=1$, but it is by no means clear that
the trend continues for higher-order bifurcation points.  For the rest
of this section, we therefore restrict attention to the case $N=1$,
that is, we take the nonlinear field $A(x,t)$ to be a one-soliton
solution of KdV (\ref{eq:kdv}).

Using (\ref{eq:simplesoliton}) and the change of variables 
$x'=\eta (x-4\eta^2 t - \alpha)$ and $t'=\eta^3 t$,
the equation (\ref{eq:linear}) becomes, after dropping primes,
\begin{equation}
\partial_t B +\partial_x [-4B + 12\kappa\,{\rm sech}^2(x)B +
\partial_x^2 B] = 0\,.
\end{equation}
This equation is of course solved by separation of variables.  We seek
separated solutions $B(x,t)=b_\sigma(x)\exp(\sigma t)$ and obtain the
third-order eigenvalue problem
\begin{equation}
[4b_\sigma(x) - 12\kappa\,{\rm sech}^2(x)b_\sigma(x) -
b_\sigma^{\prime\prime}(x)]' = \sigma b_\sigma(x)\,,
\label{eq:thirdorder}
\end{equation}
where the prime denotes differentiation with respect to $x$.  In this
context, what we have been calling ``trapped linear waves'' correspond
to bound-state eigenfunctions of (\ref{eq:thirdorder}) with
$\sigma=0$.  Such solutions have finite mass and energy and are
stationary in the moving frame of reference of the soliton $A(x,t)$.
As we know, such eigenfunctions with $\sigma=0$ exist only at the
bifurcation values of $\kappa=\kappa_n^{\rm bif}$.  However, it is
clear that for general values of $\kappa$ there are other
possibilities.  There may be eigenvalues $\sigma$ that are purely
imaginary, giving rise to oscillating modes that travel in the soliton
frame.  More generally, if an eigenvalue has a nonzero real part for
some $\kappa$, then there will be a mode that is either amplified or
exponentially damped as it propagates with the soliton.

The eigenvalue problem (\ref{eq:thirdorder}) has two simple
symmetries.  Whenever $b_\sigma(x)$ is an eigenfunction with
eigenvalue $\sigma$, then $b_\sigma(-x)$ is an eigenfunction with
eigenvalue $-\sigma$ and $b_\sigma(x)^*$ is an eigenfunction with
eigenvalue $\sigma^*$.  Therefore, the eigenvalues either come in
purely real pairs $(|\sigma|,-|\sigma|)$, purely imaginary pairs
$(i|\sigma|,-i|\sigma|)$, or in complex quartets
$(\sigma,-\sigma,\sigma^*,-\sigma^*)$.  These symmetries indicate the
distinguished role of $\sigma=0$ as a point that if it appears in the
spectrum for some $\kappa$ can signal a bifurcation in the number of
eigenvalues.  This explains our terminology and notation for the
values $\kappa=\kappa_n^{\rm bif}$.

Most points on the imaginary $\sigma$ axis correspond to continuous
spectrum.  This can be seen by the following argument.  Let $\kappa$
be fixed.  Suppose $\sigma=i\omega$ with $\omega\in {\mathbb R}$.  For
large $|x|$, the solutions of (\ref{eq:thirdorder}) have the form of
linear combinations of $\exp(ik_{\omega,j} x)$ where $k=k_{\omega,j}$
are the three roots of $k^3 + 4k -\omega = 0$.  Exactly one of these
roots, say $k=k_{\omega,0}$, is real, while the other two form a
complex-conjugate pair.  If we seek a generalized eigenfunction
normalized to $\exp(ik_{\omega,0}x)$ as $x\rightarrow -\infty$ through
a ``shooting'' method, we have three complex constants to exploit: the
coefficient of the decaying mode for large negative $x$, and the
coefficients of the decaying mode for large positive $x$ and the
finite amplitude contribution for large positive $x$.  Matching the
values of $b_{i\omega}(x)$, $b_{i\omega}'(x)$, and $b_{i\omega}''(x)$
at $x=0$ gives three complex equations in three complex unknowns.  If
this system of equations is solvable at all, one expects it to be
solvable for almost all real $\omega$, yielding a generalized
eigenfunction.  For exceptional values of $\omega$ where there is not
a generalized eigenfunction, there will be a genuine bound-state
eigenfunction, since the spectrum is a closed set.

\begin{figure}[h]
\begin{center}
 \mbox{\psfig{file=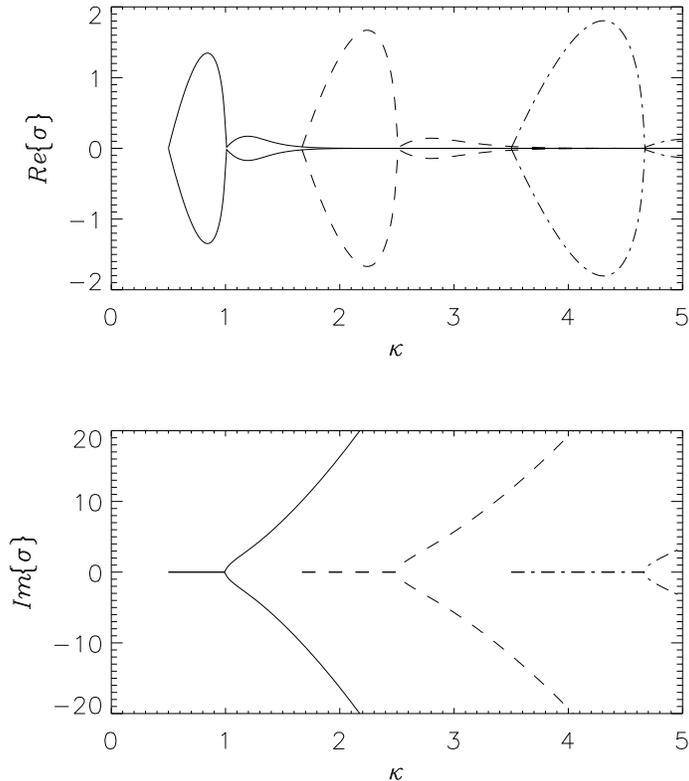,width=4 in}}
\end{center}
\caption{\em Real (above) and imaginary (below) parts of the discrete
eigenvalues $\sigma$ for the eigenvalue problem (\ref{eq:thirdorder})
as a function of the parameter $\kappa$.  Different eigenvalue
branches are displayed with different styles of lines (solid, dashed,
etc.).}
\label{fig:sigma_vs_kappa}
\end{figure}

\begin{figure}[h]
\begin{center}
 \mbox{\psfig{file=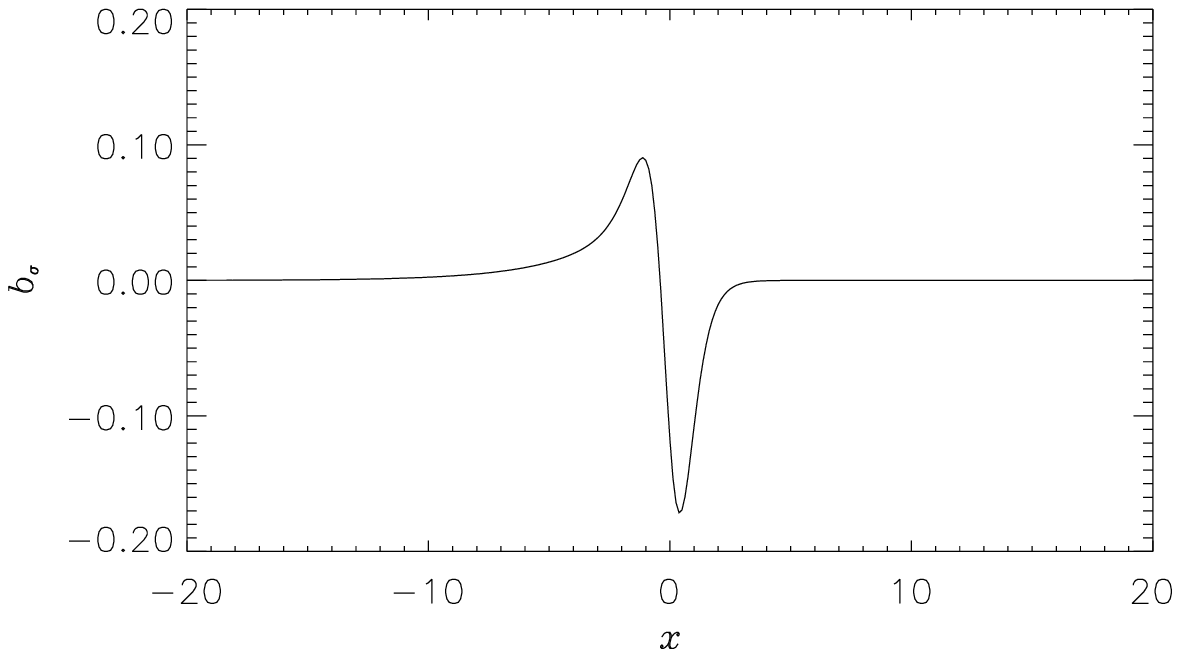,width=3 in}}
 \mbox{\psfig{file=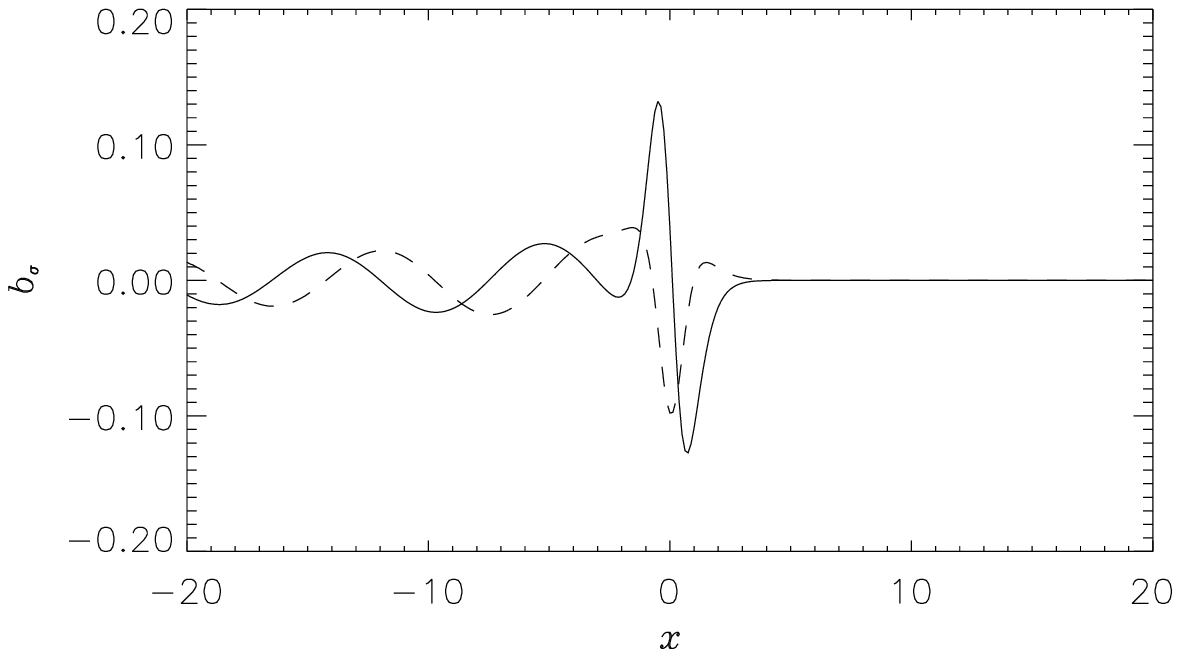,width=3 in}}
\end{center}
\caption{\em Left: the real-valued eigenfunction corresponding to the
eigenvalue $\sigma$ with positive real part for $\kappa=0.85$.  Right:
the complex-valued eigenfunction corresponding to the eigenvalue
$\sigma$ in the first quadrant for $\kappa=1.2$.  The solid curve is
the real part and the dashed curve is the imaginary part.}
\label{fig:eigenfunctions}
\end{figure}

We have used a numerical Fourier-based collocation (pseudospectral)
method to find the discrete eigenvalues of (\ref{eq:thirdorder}) over
a range of values of the coupling constant $\kappa$.  Essentially this
involves approximating the continuous function $b_\sigma(x)$ by a
periodic discrete series. Then the derivative $\partial_x$ can be
approximated to exponential accuracy by a derivative matrix $\bf D$,
for which an explicit formula is given in \cite{CHQZ88}. This is then
used to construct a discrete approximation to the operator on the
left-hand side of (\ref{eq:thirdorder}). Standard techniques can then
be used to obtain the eigenvalues and eigenvectors of this matrix. The
corresponding eigenfunctions always decay exponentially, but sometimes
they decay {\em very} slowly for large $x$ of one or the other sign
--- luckily not both.  To obtain accurate results it was necessary
in these cases to change the dependent variable by multiplying by an
appropriate exponential function of $x$ to enhance the decay on the
slowly-decaying side without changing decay into growth on the other
side.  Our results over the range $0<\kappa< 5$ are shown in
Figure~\ref{fig:sigma_vs_kappa}.
The bifurcation values $\kappa^{\rm bif}_n$ appear to be of two
different types.  If $n$ is odd, then when $\kappa$ increases through
the value $\kappa=\kappa_n^{\rm bif}$, a new pair of real eigenvalues
is born from the origin $\sigma=0$.  As $\kappa$ is further increased,
the pair of eigenvalues moves at first away from the imaginary axis
and then changes direction and contracts toward the origin.  When
$\kappa$ increases through the even bifurcation value
$\kappa=\kappa_{n+1}^{\rm bif}$, the pair enters the origin and
re-emerges as a complex eigenvalue quartet.  Further increasing the
value of $\kappa$ causes the quartet of eigenvalues to move through a
maximum in the magnitude of the real part and then toward the
imaginary axis with the magnitude of the real part decreasing to zero
while the magnitude of the imaginary part increases without bound.  It
does not appear that the quartet of eigenvalues ever disappears into
the continuous spectrum, although it comes arbitrarily close as
$\kappa$ increases.  This scenario is repeated again and again as
$\kappa$ increases through each odd bifurcation value.  Representative
eigenfunctions are plotted in Figure~\ref{fig:eigenfunctions}.
Here, one can see that when the eigenvalue $\sigma$ has a nonzero
imaginary part, the decay of the eigenfunction can be quite slow on
the ``downstream'' side of the soliton.  As remarked above, this
effect is compensated for in our numerics by working with a modified
eigenfunction.

There are no discrete eigenvalues at all for $\kappa< 1/2$ (and in
particular for $\kappa < 0$), and for all $\kappa$ satisfying
$1/2<\kappa < 1$ and $\kappa>1$, there is always at least one
eigenvalue with a nonzero real part, which corresponds to an
exponentially growing eigenfunction and therefore instability.  The
values $\kappa=1/2$ and $\kappa=1$ are distinguished as the only
values for which there exist discrete eigenvalues and at the same time
all eigenvalues have zero real parts, so the system is neutrally
stable.  For all other values of $\kappa$, either there are no
discrete eigenvalues at all in which case all initial conditions for
(\ref{eq:linear}) disperse away algebraically in time, or there are
discrete eigenvalues with positive real parts in which case the linear
waves are amplified by the soliton in the field $A(x,t)$.

\section{Conclusion}
The coupled system consisting of the KdV equation (\ref{eq:kdv}) and
the linear equation (\ref{eq:linear}) is integrable for two distinct
values of the coupling parameter $\kappa$.  The integrable case of
$\kappa=1$ has been studied by other authors \cite{GGKM74,S83}.  In
this paper, we have given new results for the other integrable case,
namely $\kappa=1/2$.  In particular, we have shown how to construct
the general solution of (\ref{eq:linear}) for $\kappa=1/2$ when the
nonlinear field $A(x,t)$ is an $N$-soliton solution of the KdV
equation.  This general solution is represented in terms of a number
of bound states (equal to the number $N$ of solitons in the field
$A(x,t)$) and a continuum superposition of radiative states given by a
singular integral.  With the help of numerical computations, we have
placed the integrable cases in context by examining the behavior of
the linear equation (\ref{eq:linear}) for general values of $\kappa$,
when $A(x,t)$ is a one-soliton solution of KdV.  These calculations
show that the linear equation (\ref{eq:linear}) behaves as an unstable
dynamical system for most positive $\kappa$.  The integrable value of
$\kappa=1$, for which the equation (\ref{eq:linear}) is the linearized
KdV equation, is an isolated stable point, since a small change of
either sign in the value of $\kappa$ will lead to the presence of
exponentially growing modes.  The other integrable value of
$\kappa=1/2$ represents the boundary between a stable system without
any bound states for $\kappa< 1/2$ and an exponentially unstable
system for $\kappa> 1/2$.

In physical applications of the coupled system (\ref{eq:kdv}) and
(\ref{eq:linear}) as discussed in the Appendix, the presence of
instabilities indicates that more terms need to be included in the
model.  However, in the stable cases the model is indeed expected to
be physically meaningful.  And in this regard, the two integrable
cases can provide useful starting points for perturbation theory.

As a final remark, let us indicate the kind of calculations that are
possible for the coupled system (\ref{eq:kdv}) and (\ref{eq:linear})
for $\kappa=1/2$ with the aid of the completeness relation.  For a
family of relevant initial data for the linear equation, one can
explicitly compute the projection onto the bound states, and
consequently determine the long time behavior of the corresponding
solution of (\ref{eq:linear}).  Also, the long time behavior of the
dispersive part of the solution can be computed from the explicit
representation of this component of the solution as a singular
integral.  We leave such applications of the completeness relation
for further investigations.

\appendix \renewcommand{\thesection}{Appendix:}
\section{Some Applications}
It is useful to keep in mind some applications in which the coupled
system (\ref{eq:kdv}) and (\ref{eq:linear}) might arise.  In fact,
such equations appear in the modeling of coupling of acoustic phonons
in long polymer molecules.  Many organic polymers ({\em e.g.} DNA and
$\alpha$-helix proteins like acetanelide) may be considered from the
mechanical point of view as long chains of nearly identical masses.
This ``backbone'' of the molecule supports a longitudinal vibrational
mode in which the masses are all moving in tandem with zero frequency
({\em i.e.} simple translation) in the long-wave limit; the associated
quanta are called acoustic phonons.  In the presence of intrinsic weak
nonlinearity, the Korteweg-de Vries equation describes these
vibrations in the long-wave limit.  Usually, the masses making up the
chain contain internal degrees of freedom ({\em e.g.}  the ``breathing''
modes of base-pairs in DNA, and the so-called amide I exciton modes of
the C=O bond in each peptide group of an $\alpha$-helix protein).  The
coupling of these internal degrees of freedom to the motion of the
backbone leads to a variety of interesting dynamical models ({\em e.g.}  the
discrete sine-Gordon equation for DNA, and the discrete nonlinear
Schr\"odinger equation for $\alpha$-helix proteins).

\begin{figure}[h]
\begin{center}
 \mbox{\psfig{file=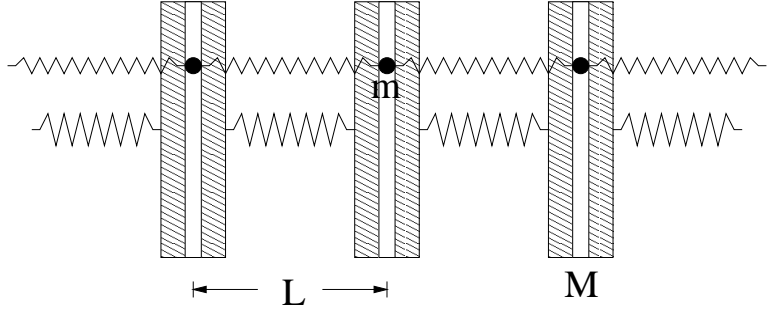,width=4 in}}
\end{center}
\caption{\em The equilibrium configuration of the mechanical model.}
\label{fig:coupled_phonons}
\end{figure}

We may consider a situation in which the internal degrees of freedom
are themselves acoustic phonons associated with transverse vibrational
modes.  This can be visualized with the help of a concrete mechanical
model, whose equilibrium configuration is shown in
Figure~\ref{fig:coupled_phonons}.
The backbone is made of heavy masses $M$ connected by stiff springs.
Mounted on each heavy mass is a transversely-oriented frictionless
track in which rides a small mass $m$. The mass $M$ is assumed to
include the mass of the track and small mass $m$.  The small masses
are themselves connected by weaker springs.  Assigning longitudinal
displacements $u_n$ to the large masses $M$, and transverse
displacements $v_n$ to the small masses $m$ in the frictionless
tracks, the Hamiltonian of the mechanical model is:
\begin{equation}
H=\sum_n \left[\frac{1}{2}M\dot{u}_n^2 +\frac{1}{2}m\dot{v}_n^2 +
W(L+u_{n+1}-u_n) + V\left(\sqrt{(L+u_{n+1}^2-u_n)^2 + (v_{n+1}-v_n)^2}
\right)\right]\,,
\end{equation}
where $W$ is the potential energy of the stiff springs connecting the large
masses and $V$ is the potential energy of the weaker springs.

The associated equations of motion are
\begin{equation}
\begin{array}{rcl}
M\ddot{u}_n &=& W'(L+u_{n+1}-u_n)-W'(L+u_n-u_{n-1})\\\\
&&\displaystyle\,\,+\,\,S(D_{n+1})(L+u_{n+1}-u_n)-
S(D_n)(L+u_n-u_{n-1})\\\\
m\ddot{v}_n&=& 
\displaystyle
S(D_{n+1})(v_{n+1}-v_n)-
S(D_n)(v_n-v_{n-1})\,,
\end{array}
\end{equation}
where we have set $S(D):= V'(D)/D$ and $D_n:=
\sqrt{(L+u_n-u_{n-1})^2+ (v_n-v_{n-1})^2}$.  It is clear that one may
take the undisturbed state of the internal modes $v_{n+1}=v_n$ for all
$n$ to hold exactly, in which case only the backbone motion is
relevant.  We will be interested in small amplitude, linear motions of
the $v_n$ and how they are affected by the motion of the backbone.

The disparity between the masses $M$ and $m$, and that of the
strengths of the associated springs, is introduced by letting $\mu$ be
a small parameter and assuming $m=\mu M$ and $V=\mu U$ (and
correspondingly, $S=\mu Z$).  We make the small-amplitude long-wave ansatz:
\begin{equation}
u_n(t)=h u(X=hn,T=ht)\,,\hspace{0.2 in}\mbox{ and }\hspace{0.2 in}
 v_n(t)=hv(X=hn,T=ht)\,,
\end{equation}
where $h$ is a small lattice-spacing parameter.  Expanding the
functions $W$ and $Z$ in Taylor series about the equilibrium position,
the equations of motion become
\begin{equation}
\begin{array}{rcl}
Mh^3 \partial_T^2u &=&\displaystyle
 h^3W''(L)\partial_X^2u+ \frac{h^5}{12} W''(L)\partial_X^4u + h^5 W'''(L)\partial_Xu\cdot\partial_X^2u +
O(h^6)+O(h^3\mu)\,,\\\\
Mh^3 \partial_T^2v&=& \displaystyle
h^3Z(L)\partial_X^2v+\frac{h^5}{12} Z(L)\partial_X^4v + h^5 Z'(L)\partial_X^2u\cdot\partial_Xv+h^5 Z'(L)
\partial_Xu\cdot\partial_X^2v + O(h^6)\,.
\end{array}
\end{equation}
Trapping of $v$-waves by $u$-waves becomes possible if the wave speeds
are equal.  Therefore, we assume that $W''(L)=Z(L) = Mc^2$.  Changing
variables to $\chi=X-cT$ and $\tau=h^2 T$ yields
\begin{equation}
\begin{array}{rcl}
Mh^2 \partial_{\tau}^2u - 2Mc\partial_\chi\partial_\tau u&=&
\displaystyle \frac{Mc^2}{12}\partial_\chi^4 u+W'''(L)\partial_\chi u\cdot
\partial_\chi^2 u
+ O(h) + O(\mu/h^2)\,,\\\\
Mh^2 \partial_\tau^2 v-2Mc\partial_\chi\partial_\tau v&=&
\displaystyle
\frac{Mc^2}{12}\partial_\chi^4 v+Z'(L)\partial_\chi^2 u\cdot\partial_\chi v
+Z'(L)\partial_\chi u\cdot\partial_\chi^2 v+O(h)\,.
\end{array}
\end{equation}
As $h\downarrow 0$ with $\mu\ll h^2$, we find the coupled system
\begin{equation}
\begin{array}{rcccl}
\partial_\tau A &+&\displaystyle \partial_\chi
\left[\frac{1}{2}A^2+\frac{c}{24}\partial_\chi^2 A\right]&=&0\,,\\\\
\partial_\tau B &+&\displaystyle \partial_\chi
\left[\frac{Z'(L)}{W'''(L)}AB + \frac{c}{24}\partial_\chi^2 B\right]&=&0\,,
\end{array}
\end{equation}
as a formal limit, where $A=W'''(L)\partial_\chi u/(2Mc)$ and
$B=\partial_\chi v$.  After a simple rescaling of $\chi$ and $\tau$,
this becomes (\ref{eq:kdv}) and (\ref{eq:linear}) with
$\kappa=Z'(L)/W'''(L)$.  As described in \S~\ref{sec:numerics}, the
influence of solitons on linear waves can be qualitatively different
for different values of the coupling constant $\kappa$, with important
bifurcations occurring at the values $\kappa=\kappa^{\rm bif}_n =
(n+1)(n+2)/12$, for $n=1,2,3,\dots$.  As we have seen, this coupled
system can be solved exactly in (at least) two cases: $\kappa=1$, and
$\kappa=1/2$.  The former case is just the linearized KdV; see Sachs
\cite{S83}.  The latter case is the one that is solved in the main
text of this paper.

Consider this example with the potential of the strong and weak springs
given respectively by
\begin{equation}
  W(D) := \frac{1}{2} \kappa_w D^2 + \frac{1}{24}\alpha D^4\,, \hspace{0.2 in}
\mbox{ and }\hspace{0.2 in}
  V(D) := \mu(\frac{1}{2} \kappa_v D^2 + \frac{1}{24}\beta D^4)\,.
\end{equation}
Thus $\beta = 3\alpha\kappa$ and the condition that the wave speeds are
equal is
\begin{equation}
  \frac{1}{2} \alpha L^2 (\kappa-1) = \kappa_w-\kappa_v\,.
\end{equation}
The effect of each bifurcation point in $\kappa$ is now clear.  At
$\kappa = 1/2$ the first harmonic of the $v$-waves begins to resonate
with the $u$-waves. As $\kappa$ increases through $\kappa=1$ we
pass through a transition from supercritical resonance to
subcritical resonance. Similarly, at the odd bifurcation points,
$\kappa=\kappa^{\rm bif}_{2m-1}$, for $m=1,2,3,\dots$, the $m^{\rm
th}$ harmonic $v$-wave begins to resonate with the $u$-waves.  Then at
the even bifurcation points, $\kappa=\kappa^{\rm bif}_{2m}$, for
$m=1,2,3,\dots$, the nature of the resonance for this mode changes
from supercritical to subcritical.

Coupled systems of equations like the pair
(\ref{eq:kdv}) and (\ref{eq:linear}) often arise as formal asymptotic
reductions of mechanical models for complicated one-dimensional waves.
Often these asymptotic models are integrable.  For example, in an
elastic rod, the interaction between axial twist waves and helical
deformation waves gives rise to an integrable Manakov system of
coupled nonlinear Schr\"odinger equations \cite{LG99}.

The coupled system (\ref{eq:kdv}) and (\ref{eq:linear}) for
$\kappa=1/2$ is also intimately connected with an integrable
multicomponent (an arbitrary number of components, all appearing
symmetrically) coupled KdV equation \cite{MC99}.  Indeed, from one
point of view it is this connection that yields the solvability of
(\ref{eq:kdv}) and (\ref{eq:linear}) for $\kappa=1/2$ described in
detail in this paper.  The solution method presented here also can be
used to give the complete solution of the coupled KdV system.  That
system in turn can be interpreted as a phenomenological model for the
transport of the mass integral through an $N$-soliton solution of KdV
\cite{MC99}.

Finally, we would like to point out that there are also some
applications in which linear equations of the form (\ref{eq:linear})
occur with $\kappa A(x,t)$ being a given function.  In this case,
$c(x,t)=\kappa A(x,t)$ represents a given spatiotemporal modulation of
the speed of linear dispersive waves, say due to propagation in an
inhomogeneous medium.  Such problems arise in the modeling of the
propagation of weak internal waves in a channel of varying width
\cite{CG99}.  For such applications, we may view the solvability of
the coupled system (\ref{eq:kdv}) and (\ref{eq:linear}) for $\kappa=1$
and $\kappa=1/2$ as a kind of (big) catalog of special cases of the
function $c(x,t)$ for which the linear equation (\ref{eq:linear}) is
solvable in its own right.  For other values of $\kappa > 1/2$ the
linear wave system is unstable, while for all values of $\kappa< 1/2$
it is stable.

\end{document}